# Derivative coordinates for analytic tree fractals and fractal engineering

Henk Mulder

henk.mulder@geneticfractals.com

**Abstract**

We introduce an alternative coordinate system based on derivative polar and spherical coordinate functions and construct a root-to-canopy analytic formulation for tree fractals. We develop smooth tree fractals and demonstrate the equivalence of their canopies with iterative straight lined tree fractals. We then consider implementation and application of the analytic formulation from a computational perspective. Finally we formulate the basis for concatenation and composition of fractal trees as a basis for fractal engineering of which we provide some examples.

## 1. Introduction

Tree fractals are objects of interest and fascination but surprisingly, we have seen only limited application of tree fractals as foundations for systems and engineering. One notable exception are L-systems as a model for botanical structures such as plants and trees [1]. The preponderance of such fractals in nature [2] would suggest that as scientists and engineers we are missing an opportunity for efficient system design.

One of the limitations of tree fractals is in their construction. In one form or another, they are mostly iteratively constructed, i.e. the systematic branching and stacking of geometric primitives such as straight line segments [3]. Such composed structures hamper our ability to analyze them effectively. What if we had an analytical function that represented an entire tree fractal?

This question has motived the research presented here and has resulted in exactly such a function. In this paper we present an analytic formulation of tree fractals and a foundation of derivative coordinates that underpins them. We then use that formulation to derive some initial results including a formal description of fractal engineering.

## 2. Derivative coordinate functions

**Theorem 1** *Let a parametric path $p(s)$ for $p: \mathbb{R} \mapsto \mathbb{R}^2$ be defined by a pair of Cartesian coordinate functions $(x(s), y(s))$ for $x, y: \mathbb{R} \mapsto \mathbb{R}$ then $p$ may be expressed as a function of a pair of derivative coordinate functions $\left(\frac{\partial r}{\partial s}, \frac{\partial \varphi}{\partial s}\right)$ for $r, \varphi: \mathbb{R} \mapsto \mathbb{R}$, that represent the relative polar coordinates at a point $(x, y)$ on $p$. The path $p$ is then given by:*

$$p(s) = \begin{pmatrix} x(s) \\ y(s) \end{pmatrix} = \begin{pmatrix} \int \frac{\partial r}{\partial s} \cos\left(\int \frac{\partial \varphi}{\partial s} ds\right) ds \\ \int \frac{\partial r}{\partial s} \sin\left(\int \frac{\partial \varphi}{\partial s} ds\right) ds \end{pmatrix} \quad (1)$$

and $\left(\frac{\partial r}{\partial s}, \frac{\partial \varphi}{\partial s}\right)$ are defined as

$$\begin{pmatrix} \frac{\partial r}{\partial s} \\ \frac{\partial \varphi}{\partial s} \end{pmatrix} = \begin{pmatrix} \sqrt{\left(\frac{\partial x}{\partial s}\right)^2 + \left(\frac{\partial y}{\partial s}\right)^2} \\ \frac{\partial \tan^{-1} \frac{\partial y}{\partial s}/\frac{\partial x}{\partial s}}{\partial s} \end{pmatrix}. \quad (2)$$

*Notation.* In this paper we will see a lot of partial derivatives and to facilitate in line equations, from here on we will represent these with the del operator $\nabla$ and we remind ourselves that unless otherwise indicated, these are all 1-dimensional derivatives with respect to s, i.e. $\nabla r = \frac{\partial r}{\partial s}$ where $r: \mathbb{R} \mapsto \mathbb{R}$. So (1) becomes

$$p(s) = \begin{pmatrix} x(s) \\ y(s) \end{pmatrix} = \begin{pmatrix} \int \nabla r \cos(\int \nabla \varphi \, ds) \, ds \\ \int \nabla r \sin(\int \nabla \varphi \, ds) \, ds \end{pmatrix}$$

(3)

and (2)

$$\begin{pmatrix} \nabla r \\ \nabla \varphi \end{pmatrix} = \begin{pmatrix} \sqrt{(\nabla x)^2 + (\nabla y)^2} \\ \nabla \tan^{-1} \nabla y / \nabla x \end{pmatrix}$$

(4)

We will refer to $\nabla r$ as the radial derivative coordinate and to $\nabla \varphi$ as the angular derivative coordinate. We may refer to the derivative coordinates functions $(\nabla r, \nabla \varphi)$ simply as derivative coordinates, suggesting derivative coordinate values for a specific value of the parametric variable $s$.

We will favor the Euler form of (3) for most of this paper,

$$p(s) = x(s) + iy(s) = \int \nabla r \exp(\int i \nabla \varphi \, ds) \, ds$$

(5)

but when developing the n-dimensional forms where n > 2, we will revert to the trigonometric form.

*Proof.* Substitute (4) into (3), noting that the integrals cancel out the derivatives with respect to the same variable s, we find that $\begin{pmatrix} x(s) \\ y(s) \end{pmatrix} = \begin{pmatrix} x(s) \\ y(s) \end{pmatrix}$. □

The derivative coordinates $(\nabla r, \nabla \varphi)$ are purposefully written as derivatives since geometrically they describe how a path evolves with respect to a point on the path, i.e. the relative radial and angular evolution. For that reason we do not evaluate the integral $\int \nabla \varphi \, ds = \varphi$ to remind ourselves that $\nabla \varphi$ is our coordinate, and not $\varphi$.

However, there are two cases when these integrals are interesting in their own right.

***Corollary 1.1*** *Let a parametric path $p(s)$ be represented by derivative coordinates functions $(\nabla r, \nabla \varphi)$ and the path equation $\int \nabla r \exp(\int i \nabla \varphi \, ds) \, ds$, then the path length l between two points at $s_1$ and $s_2$ is*

$$l = \int_{s_1}^{s_2} \nabla r \, ds. \tag{6}$$

*Proof.* The derivative coordinate $\nabla r$ in (4) is defined as $\nabla r = \sqrt{(\nabla x)^2 + (\nabla y)^2}$, i.e. the differential arc length of $p$ at point $(x(s), y(s))$, it follows that the arc length from $s_1$ and $s_2$ is $\int_{s_1}^{s_2} \nabla r \, ds$. □

***Corollary 1.2*** *Let a parametric path $p(s)$ be represented by derivative coordinates functions $(\nabla r, \nabla \varphi)$ and the path equation $\int \nabla r \exp(\int i \nabla \varphi \, ds) \, ds$, then the absolute angle $\Delta \Phi$ between the tangent lines at two points at $s_1$ and $s_2$ is*

$$\Delta \Phi = \int_{s_1}^{s_2} \nabla \varphi \, ds \tag{7}$$

*Proof.* From the definition (2) we note that $\int \nabla \varphi \, ds = \tan^{-1} \frac{\partial y}{\partial s} / \frac{\partial x}{\partial s}$. By eliminating s we find the anti-derivative $\Phi = \tan^{-1} \frac{\partial y}{\partial x}$ which is the angle of the tangent at $s$ on the path $p$ with respect to x-axis. It follows that $\int_{s_1}^{s_2} \nabla \varphi \, ds = \Phi_2 - \Phi_1$, i.e. the absolute angle $\Delta \Phi$ between the tangent lines at points $s_1$ and $s_2$. □

We will extend the path formulation of Theorem 1 to the third dimension.

***Corollary 1.3.*** *Let a parametric path $p(s)$ for $p: \mathbb{R} \mapsto \mathbb{R}^3$ be defined by the Cartesian coordinate functions $(x(s), y(s), z(s))$ for $x, y, z: \mathbb{R} \mapsto \mathbb{R}$ then $p$ may be expressed as a function of a triple of derivative coordinate functions, $(\nabla r(s), \nabla \varphi(s), \nabla \psi(s))$ for $r, \varphi, \psi: \mathbb{R} \mapsto \mathbb{R}$ that represent the relative spherical coordinates at a point $(x, y, z)$ on $p$. The path $p$ is then given by:*

$$p(s) = \begin{pmatrix} x(s) \\ y(s) \\ z(s) \end{pmatrix} = \begin{pmatrix} \int \nabla r \cos(\int \nabla \varphi \, ds) \sin(\int \nabla \psi \, ds) \, ds \\ \int \nabla r \sin(\int \nabla \varphi \, ds) \sin(\int \nabla \psi \, ds) \, ds \\ \int \nabla r \cos(\int \nabla \psi \, ds) \, ds \end{pmatrix} \tag{8}$$

*and $(\nabla r(s), \nabla \varphi(s), \nabla \psi(s))$ are defined as*

$$\begin{pmatrix} \nabla r \\ \nabla \varphi \\ \nabla \psi \end{pmatrix} = \begin{pmatrix} \sqrt{(\nabla x)^2 + (\nabla y)^2 + (\nabla z)^2} \\ \nabla \tan^{-1} \nabla y / \nabla x \\ \nabla \cos^{-1} \nabla z / \sqrt{(\nabla x)^2 + (\nabla y)^2 + (\nabla z)^2} \end{pmatrix} \qquad (9)$$

*Proof.* (9) and (10) are the standard spherical coordinate system. Substitution of (9) into (10) verifies that both spherical and Cartesian forms are equal. □

More generally, we can extend this to any higher dimension.

***Corollary 1.4.*** *Let a parametric path $p(s)$, $p: \mathbb{R} \mapsto \mathbb{R}^n$ for $n > 3$ be defined by Cartesian coordinate functions $(x_i(s) | 4 \leq i \leq n, i \in \mathbb{N})$ where $x_i: \mathbb{R} \mapsto \mathbb{R}$ then $p$ may be expressed as a function of derivative coordinate functions, $(\nabla v_i | 4 \leq i \leq n, i \in \mathbb{N})$ where $v_i: \mathbb{R} \mapsto \mathbb{R}$ represent the relative spherical coordinates on an n-sphere at a point $(x_i | 4 \leq i \leq n, i \in \mathbb{N})$ on p.*

*Proof.* We will not give a formal proof but note that as long as the relative evolution of a path $p(s)$, locally within a hyper spherical coordinate system is unrestricted, then the integral of the relative evolution will describe any path $p(s)$ for $p: \mathbb{R} \mapsto \mathbb{R}^n$ □

## 3. Application of derivative coordinate functions in tree fractals

This research was initiated with the purpose of finding an analytic formulation of tree fractals. Before we can demonstrate the use of derivative coordinate functions, we need to introduce a formulation for branches in tree fractals.

If we accept, or define it to be as such, that $\sqrt{x^2}$ has two solutions $(x, -x)$ for $x \neq 0$ and one solution (0) for $x = 0$, then $f(x) = \sqrt{x^2}$ is a multivalued function with a node point at $x = 0$ where $f(x)$ is single valued. Taking this one step further, a function $\sqrt{\sin^2(x)}$ is a multivalued function with node points at $x = n\pi$ for $n \in \mathbb{Z}$ where $f(x)$ is single valued. When we integrate such a function, each time we get to a node where f(x) is single valued, the function becomes dual valued again, i.e. it forks. This is repeated at every node point and the repetitive forking creates a multivalued tree. This multivalued integral $\int \sqrt{\sin^2(x)} dx$ is shown below.

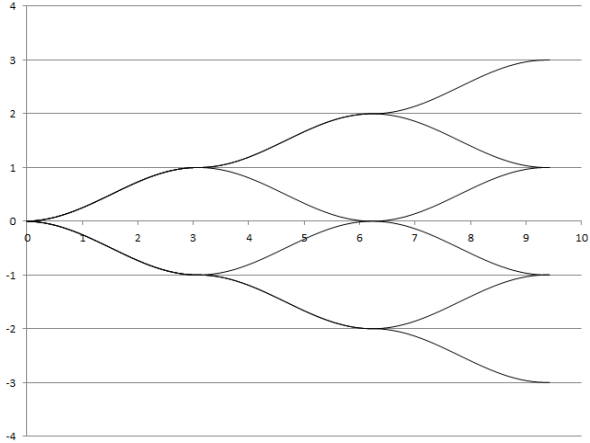 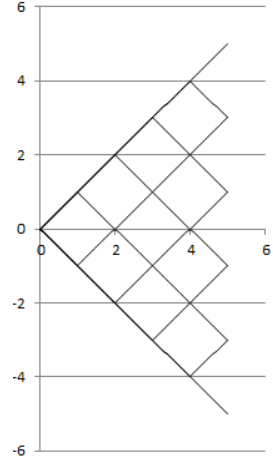

Image 1: multivalued integral of $\sqrt{\sin^2(x)}$     Image 2: multivalued integral of $\check{u}(s)$

We could use such an analytic multivalued integral as a mechanism for tree branching however, we prefer to define a function that is linear.

***Definition 2***. *We define a multivalued periodic unit function $\check{u}(s)$ such that $\check{u}(s) = \begin{cases} 1, -1 & s \notin B \\ 0 & s \in B \end{cases}$ where B is a branch point set, $B = \{s_i | \forall s_i : i \in \mathbb{N}, s \in \mathbb{R}\}$.*         *(9)*

The multivalued integral of the multivalued periodic unit function is shown in image 2.

***Theorem 2***. *Using suitable derivative functional coordinates where the angular coordinate $\nabla\varphi$ is multiplied by the multivalued periodic unit function $\check{u}(s)$ with a branch point set $B = \{s_i | \forall s_i : i \in \mathbb{N}, s \in \mathbb{R}\}$, then any symmetric binary tree is described by the function*

$$\check{p}(s) = \int \nabla r \exp(\int i\check{u}\nabla\varphi \, ds) \, ds \qquad (10)$$

*Notation*. The "check" symbol over $\check{p}$ indicates that this is a tree function. It may be pronounced as *p-tree*.

*Proof*. We will demonstrate that the tree is symmetric, binary and that the formulation can represent any tree.

At each branch node $\check{p}(s_i)$ where $s_i \in B$, the direction of the two emanating branches is determined uniquely by the term $\check{u}\nabla\varphi$ which has the multi-values $\nabla\varphi$ and $-\nabla\varphi$ in between the branch points such that the respective branch paths $p_a$ and $p_b$ are given by:

$$p(s) = \int \nabla r \exp(\int i\nabla\varphi \, ds) \, ds \qquad (11)$$

and

$$p(s) = \int \nabla r \exp(-\int i\nabla\varphi \, ds) \, ds \tag{12}$$

We note that the only difference between these paths $p_a$ and $p_b$ is in the direction of the rotation $\int \nabla\varphi \, ds$ and $-\int \nabla\varphi \, ds$ but not in the quantity. It follows that these two branches are symmetric. Since all branches are symmetric from the branch point from which they emanate and that at each branch point there are precisely two branches, it follows that the tree is symmetric and binary. □

We note that at a branch point $s_i$, the multivalued periodic unit function is single valued, i.e. $\breve{u} = 0$ and we shall consider that case in the following corollary.

***Corollary 2.1*** *A tree defined by Theorem 2, $\breve{p}(s) = \int \nabla r \exp(\int i\breve{u}\nabla\varphi \, ds) \, ds$ is continuous at every point of its domain $s \in \mathbb{R}$, if and only if the derivative coordinates functions $(\nabla r, \nabla\varphi)$ are continuous on $r(s), \varphi(s): \mathbb{R} \mapsto \mathbb{R}$.*

*Proof.* It is clear that the tree is continuous between consecutive tree nodes since at these intervals of the domain the path function $p(s)$ reduces to the path function of Theorem 1, $p(s) = \int \nabla r \exp(\int i\nabla\varphi \, ds) \, ds$. At the branch points $s_i$, we consider the limits of the integral of the angular coordinates of the right branch $p_a$ from below and from above

$$\lim_{s \uparrow s_i} \int \nabla\varphi_a \, ds = \lim_{s \downarrow s_i} \int \nabla\varphi_a \, ds \tag{13}$$

and for the left branch $p_b$

$$\lim_{s \uparrow s_i} \int \nabla\varphi_b \, ds = \lim_{s \downarrow s_i} \int -\nabla\varphi_b \, ds \tag{14}$$

it follows that from (13) and (14)

$$p_a(s_i) = p_b(s_i) \tag{15}$$

at the branch points and thus a tree defined by Theorem 2 is continuous on its domain. □

***Corollary 2.2*** *A tree defined by Theorem 2, $\breve{p}(s) = \int \nabla r \exp(\int i\breve{u}\nabla\varphi \, ds) \, ds$ is analytic and continuous on its entire domain $s \in \mathbb{R}$ if and only if the derivative coordinates functions $(\nabla r, \nabla\varphi)$ are analytic and continuous on $r(s), \varphi(s): \mathbb{R} \mapsto \mathbb{R}$.*

*Proof.* Since by Corollary 2.1 a tree defined by Theorem 2, $\breve{p}(s) = \int \nabla r \exp(\int i\breve{u}\nabla\varphi \, ds) \, ds$ is continuous and $(\nabla r, \nabla\varphi)$ are analytic for $r(s), \varphi(s): \mathbb{R} \mapsto \mathbb{R}$, we note that we may write

$$\breve{p}(s) = f(\nabla r, \nabla\varphi) \tag{16}$$

where $f$ itself is a composite analytic function, it follows that a tree defined by Theorem 2 is analytic and continuous on $r(s), \varphi(s): \mathbb{R} \mapsto \mathbb{R}$. □

***Lemma 2.3***. *A symmetric binary tree is fully defined by any of its branch paths, from the root to any of its branch tips.*

*Proof.* Starting from root, after the first branch node, the binary branches are symmetric i.e. one defines the other and vice versa. This is the case for any two branches at a branch point. It follows that a symmetric binary tree fractal is fully defined by any of the paths that start at the root and end at a branch tip. □

*Notation.* From here on, when referring to a tree, we are referring to a binary symmetric tree unless otherwise stated.

***Corollary 2.4*** *Theorem 2 may define any tree.*

*Proof.* By Lemma 2.3 we note that a tree is defined by any of its branch paths. If we select the branch path where $\breve{u}(s) = 1$ for all its nodes, then the path $p$ is given by:

$$p(s) = \int \nabla r \exp(\int i \nabla \varphi \, ds) \, ds \tag{17}$$

By Theorem 1 such a path function may represent any path $p$, it follows that Theorem 2 may describe any tree. □

***Corollary 2.5***. *Using suitable derivative functional coordinates $(\nabla v_i | 3 \leq i \leq n, i \in \mathbb{N})$ where $v_i \colon \mathbb{R} \mapsto \mathbb{R}$ where the directional coordinates $\nabla v_i$ are multiplied by the multivalued periodic unit function $\breve{u}_d(s)$ with a given branch set $B_d = \{s_i | \forall s_i \colon i \in \mathbb{N}, s \in \mathbb{R}\}_d$ for dimension $d \geq 3$, then any tree may be described by the higher dimensional path functions defined under Corollaries 1.3 and 1.4.*

*Proof.* We will not provide a proof but note that with addition that each additional dimension $d \geq 3$ we have added a corresponding multivalued periodic unit function $\breve{u}_d(s)$ with an associated branch point set $B_d = \{s_i | \forall s_i \colon i \in \mathbb{N}, s \in \mathbb{R}\}_d$. With this addition, the argument of the proof of Theorem 2 can be shown to be valid for dimensions $d \geq 3$. □

***Lemma 2.6***. *A tree defined by Theorem 2 is bounded if the length of its branches is bounded over its domain $s \in \mathbb{R}$.*

*Proof.* A tree defined by Theorem 2 is essentially an integral of free differential polar vectors. It is clear that if the integral of the angular coordinate $\int \nabla \varphi \, ds = c$ for some constant $c$, then the path represented by $\breve{u}(s) = 1$ for all branches on this longest possible path is a straight line segment at angle $c$ with respect to a Cartesian origin point. The length $l$ of this line segment is

$$l = \int \nabla r \, ds \tag{18}$$

And the tree is bounded by a circle centered on the root with radius $l$. Therefore a tree as defined by Theorem 2 is bounded if $\int \nabla r \, ds$ is bounded. □

In some cases we can be more precise than this and calculate the smallest boundary circle explicitly.

**Lemma 2.7.** *A non-trivial tree defined by Theorem 2, i.e. where the derivative coordinates $\int \nabla r \, ds \neq 0$ and $\int \nabla \varphi \, ds \neq c$ for some constant c, has a non-integer Hausdorff dimension that exceeds its topological dimension.*

*Proof.* We do not give a formal proof but note that the canopy of this tree is constructed from infinitely many non-trivial branches within a bounded area. It follows that the canopy of this tree is infinitely intricate. One may also note that regardless of the particular shape of the branches, their distinctness will increase a box count from generation to generation when assessing the specific Haussdorf dimension for this tree. □

**Theorem 3.** *A tree as defined by Theorem 2 is an exact self-similar fractal if the derivative coordinates satisfy the property that between any two consecutive parent and child branches defined by $s_{ab} \in [a..b]$ and $s_{bc} \in [b..c]$ where a,b,c are branch points, that the scaling between branch points is constant*

$$\nabla r \, |_{s=b..c} = \rho \nabla r \, |_{s=a..b} \text{ and } 0 < \rho < 1 \tag{19}$$

and the directional change is equivalent

$$\nabla \varphi \, |_{s=b..c} = \nabla \varphi \, |_{s=a..b} \tag{20}$$

and branch points $s_i$ are equidistant

$$s_{i-1} - s_i = c \text{ for } c > 0 \tag{21}$$

*Proof.* We will follow the characteristics of fractals outlined by Falconer [4]. By Lemma 2.7, the bounded binary symmetric tree has a Hausdorff dimension that exceeds its topological dimension. The conditions that $\nabla r \, |_{s=b..c} = \rho \nabla r \, |_{s=a..b}$ and $\nabla \varphi \, |_{s=b..c} = \nabla \varphi \, |_{s=a..b}$ for consecutive parent and child branches ensures that parent and child are exactly self-similar. Finally, the definition of the tree by Theorem 2 is simple. □

**Corrolary 3.1.** *If either $\nabla r \, |_{s=b..c} = \rho \nabla r \, |_{s=a..b}$ or $\nabla \varphi \, |_{s=b..c} = \nabla \varphi \, |_{s=a..b}$, but not both, then a bounded binary symmetric tree formulated by Theorem 2 for the branch point set $B = \{s_i | \forall s_i : i \in \mathbb{N}, s \in \mathbb{R}\}$ is quasy fractal.*

*Proof.* No formal proof is given but it is clear from the condition $\nabla r \, |_{s=b..c} = \rho \nabla r \, |_{s=a..b}$ or $\nabla \varphi \, |_{s=b..c} = \nabla \varphi \, |_{s=a..b}$, that some but not all fractal features are carried from branch to branch across the generations. □

**Remark 3.2.** *The conditions of Theorem 3 and Corollary 3.1 on the self-similarity across branch generations may be relaxed by allowing the sets $s_{ab}$ and $s_{bc}$ to be defined less rigidly between branch points but across arbitrary intervals of s as long as these intervals are repeated ad infinitum. The resulting fractal will be exact self-similar.*

# 4. Computational considerations

## 4.1 Calculating curves in $\mathbb{R}^n$

The definition of the derivate coordinate functions as an alternative representation of a curve in $\mathbb{R}^n$ simplifies the calculation of visual rendering of curves and objects in $\mathbb{R}^2$ and $\mathbb{R}^3$, in particular when these curves are not available algebraically. Furthermore, the calculation of curves and objects in $\mathbb{R}^n, n > 3$ is also simplified.

We recall that rendering curves and objects in computer graphics often involves drawing a point $p_2$ at a given distance $\Delta r$ and rotation $\Delta \varphi$ with respect to some point $p_1$ such that the vector $\Delta \boldsymbol{v}(\Delta r, \Delta \varphi) = p_2 - p_1$. The usual approach is to translate ($\boldsymbol{T}$) $p_1$ back to the origin, rotate ($\boldsymbol{R}$) it such that it is parallel to the $x$-axis and sometimes we may need to scale ($\boldsymbol{S}$) it back to unit length as well. Then we add the new line segment and reverse the transformations: re-scale ($\boldsymbol{S}^{-1}$), re-rotate ($\boldsymbol{R}^{-1}$), re-translate ($\boldsymbol{T}^{-1}$).

$$p_2 = (p_1 \cdot \boldsymbol{T} \cdot \boldsymbol{R} \cdot \boldsymbol{S} + \Delta \boldsymbol{v}) \cdot \boldsymbol{S}^{-1} \cdot \boldsymbol{R}^{-1} \cdot \boldsymbol{T}^{-1} \tag{22}$$

These transformations are typically done with matrices or quaternions which in particular for higher dimensions can add a computational complexity.

A similar calculation for a curve in $\mathbb{R}^2$ using derivate coordinate functions where $\Delta \boldsymbol{v} = (\Delta r, \Delta \varphi)$, $p_1 = (x_1, y_1)$ and the direction at that point is $\Phi_1$, then using Theorem 1:

$$p_2(s) = x_2 + iy_2 = \int \nabla r \exp(\int i \nabla \varphi \, ds) \, ds \tag{23}$$

which we implement as a Riemann sum

$$p_2(s) = x_2 + iy_2 = \sum \nabla r \exp(\sum i \nabla \varphi \, \Delta s) \, \Delta s \tag{24}$$

to calculate the next point

$$p_2 = \begin{pmatrix} x_1 + \Delta r \cos(\Phi_1 + \Delta \varphi) \\ y_1 + \Delta r \sin(\Phi_1 + \Delta \varphi) \end{pmatrix}. \tag{25}$$

Arguably, all of the transformations represented by the matrices $\boldsymbol{T}, \boldsymbol{R}, \boldsymbol{S}$ are embedded in equation (25). The inverse matrices are avoided by the integrals that 'remember' the position and direction of the curve at the last point. So what is the difference? This is exactly the point: the transformations are built into the equation that describes the path that we are calculating and we only have to concern ourselves with the derivate coordinate functions, not with transformations.

## 4.2 Tree fractals

### 4.2.1 Pseudo code for generating fractals

An example of usage of an analytical formulation of a fractal by Theorem 3 is shown in the pseudo code below.

**Initialize**

1: set initial position $\boldsymbol{p} = [0,0]$
2: set initial direction $\Phi = 0$
3: set path parametric variable $s = 0$
4: set step increment $\Delta s = 1$ [1]
5: read from a file, or calculate from a function into arrays $\nabla r[s]$, $\nabla\varphi[s]$ and $\breve{u}[s]$ [2,3]
6: EvaluateFractal($s, \Phi, \boldsymbol{p}, \breve{u}_{right}$) // first right branch [4]
7: EvaluateFractal($s, \Phi, \boldsymbol{p}, \breve{u}_{left}$) // first left branch

**Function** $\Phi_{next}, \boldsymbol{p_{next}}$=FractalEquation( $\nabla r, \nabla\varphi, \boldsymbol{p_{last}}, \Phi_{last}$)

1: $\Phi_{next} = \Phi_{last} + \nabla\varphi$
2: $\begin{bmatrix} x_{last} \\ y_{last} \end{bmatrix} = p_{last}$
3: $\boldsymbol{p_{next}} = \begin{bmatrix} x_{next} \\ y_{next} \end{bmatrix} = \begin{bmatrix} x_{last} + \nabla r \cos \nabla\varphi\, \Delta s \\ y_{last} + \nabla r \sin \nabla\varphi\, \Delta s \end{bmatrix}$ // refer equation (25)
4: **return** $\Phi_{next}, \boldsymbol{p_{next}}$

**Recursive** EvaluateFractal($s, \Phi_{last}, \boldsymbol{p_{last}}, \breve{u}$)

1: **while** $s < s_{max} \wedge \breve{u}[s] \neq 0$ **do** // test for eof and branch condition
2:     $\Phi_{next}, \boldsymbol{p_{next}}$=FractalEquation( $\nabla r[s], \breve{u}\nabla\varphi[s], \boldsymbol{p_{last}}, \Phi_{last}$)
3:     PlotLine($\boldsymbol{p_{last}}, \boldsymbol{p_{next}}$) // use a canvas or write to coordinate file for offline plotting
4:     $\boldsymbol{p_{last}} = \boldsymbol{p_{next}}$
5:     $\Phi_{last} = \Phi_{next}$
6:     $s = s + \Delta s$
7: **end while**
8: **if** $s < s_{max}$ **then** // this implies that $\breve{u}(s)$ was 0
9:     EvaluateFractal($s, \Phi_{last}, \boldsymbol{p_{last}}, \breve{u}_{right}[s]$) // next right branch
10:     EvaluateFractal($s, \Phi_{last}, \boldsymbol{p_{last}}, \breve{u}_{left}[s]$) // next left branch
11: **end if**

A rudimentary Javascript implementation of this program is available [6] as well as a more evolved Ruby implementation with accessory functions (refer section 4.2.4) for color, line width and transparency [7].

---

[1] If another value for $\Delta s$ is chosen, note that $\int_a^b \nabla r\, ds$ and $\int_a^b \nabla\varphi\, ds$ for a branch from a to b, should remain invariant and $\nabla r$ and $\nabla\varphi$ will need to be scaled accordingly.

[2] The multivalued periodic unit function $\breve{u}$ is either 1 and -1, or it is 0. This may be implemented as an array with two columns $\breve{u}_{right}$ and $\breve{u}_{left}$. When $\breve{u} = 0$, we may set both $\breve{u}_{right}$ and $\breve{u}_{left}$ to zero and for the purposes of a branch test agree that we can use either $\breve{u}_{right}$ or $\breve{u}_{left}$.

[3] In practice we will often implement a branch test based on intervals of *s* rather than $\breve{u}$. In that case we define two constants $\breve{u}_{right} = 1$ and $\breve{u}_{right} = -1$.

[4] Continuous tree fractals never have a trunk; that is an artifice of straight line tree fractals where a trunk is defined as an initial lone branch.

### 4.2.2 Examples of tree fractals

The use of derivate coordinate functions in tree fractals allows us to introduce new features. We will substitute the traditional straight line segments by other line shapes, including analytic functions. For example, if we set $\nabla\varphi = c$ for some constant c such that for a branch from node $s_a$ to $s_b$, the overall angular change $\Phi = \int_a^b \nabla\varphi \, ds = \frac{\pi}{3}$ for any branch and $\nabla r = \left(\frac{2}{3}\right)^{s/\Delta s}$ and position the branch points in $s$ at equidistant intervals of length $\Delta s$, the fractal function $\check{p}(s)$ of Theorem 3 will generate the fractal below (image 3) and the derivative coordinate functions $\nabla r$ and $\nabla\varphi$ and the absolute periodic unit function $|\check{u}(s)|$ that were used to generate it (image 4)

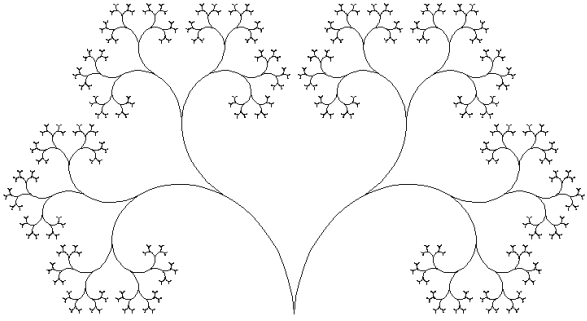
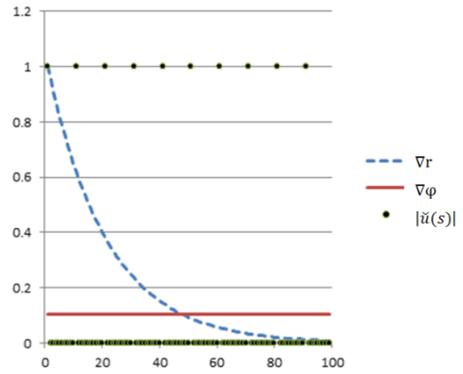

Image 3: Smooth fractal : $\Phi = \frac{\pi}{3}$ and $\nabla r = \left(\frac{2}{3}\right)^{s/\Delta s}$

Image 4: corresponding $\nabla r$, $\nabla\varphi$ and $|\check{u}(s)|$

For comparison, we generate a straight line fractal using the similar parameters that are only assigned at the branch points and zero elsewhere. So,

- set $\nabla\varphi = c_\varphi$ for some constant $c_\varphi$ at the branch points and $\nabla\varphi = 0$ elsewhere whilst ensuring that $\int_a^b \nabla\varphi \, ds = \frac{\pi}{3}$ and
- set $\nabla r = c_r \left(\frac{2}{3}\right)^{s/\Delta s}$ for some constant $c_r$ at the branch points and $\nabla r = 0$ elsewhere whilst ensuring that $\int_a^b \nabla r \, ds$ is invariant with respect to the same integral in the preceding smooth fractal (Image 3).

we will obtain the straight line fractal below that appears to have the same general form as the smooth fractal above. In Theorem 4 that follows, we will demonstrate the equivalence of their camopies under scaling, rotation and translation.

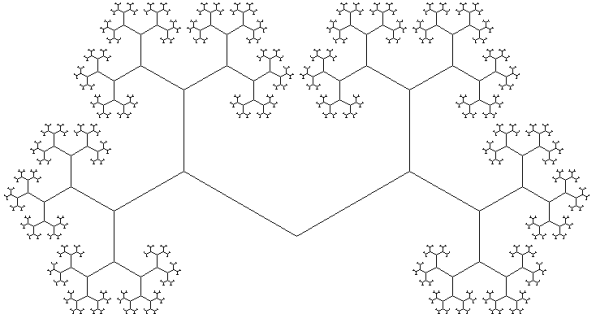

Image 5: Straight line fractal : $\Phi = \frac{\pi}{3}$ and $\nabla r = \left(\frac{2}{3}\right)^{s/\Delta s}$

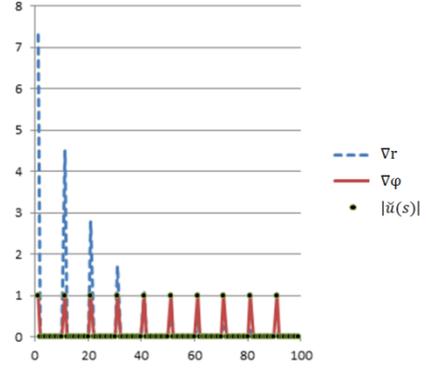

Image 6: Corresponding $\nabla r$, $\nabla \varphi$ and $|\check{u}(s)|$

It is worth pointing out that the fractals in images 3 and 5 are not the same size. The curvature of the smooth fractal in image 3 effectively shrinks the fractal compared to the straight lined fractal of image 5. The scale factor between these two fractals may be found by comparing the absolute distance of subsequent branch points $s_1$ and $s_2$ for these two fractals $\check{p}_A$ and $\check{p}_B$.

$$scale = \frac{\check{p}_A(s_1) - \check{p}_A(s_2)}{\check{p}_B(s_1) - \check{p}_B(s_2)} \qquad (26)$$

**Theorem 4.** *If two tree fractals $\check{p}_A(s)$ and $\check{p}_B(s)$ are exact self-similar, with the respective derivative coordinate functions $(\nabla r_A, \nabla \varphi_A)$ and $(\nabla r_B, \nabla \varphi_B)$ then if between any consecutive branch points $s_1$ and $s_2$ the angular integral*

$$\int_{s_1}^{s_2} \nabla \varphi_A ds = \int_{s_1}^{s_2} \nabla \varphi_B ds \qquad (27)$$

*and branch length*

$$\int_{s_1}^{s_2} \nabla r_A ds = c \int_{s_1}^{s_2} \nabla r_B \, ds \qquad (28)$$

*for some constant c then in the limit for $s \to \infty$, the canopies of these fractals are equivalent through scaling, translation and rotation.*

*Proof.* We consider two binary symmetric tree fractals with the respective derivative coordinate functions $(\nabla r_a, \nabla \varphi_a)$ and $(\nabla r_b, \nabla \varphi_b)$ then if we can demonstrate that with a suitable scaling, rotation and translation the corresponding node points between the two fractals are arbitrarily close at $s \to \infty$ then these fractals are equivalent through scaling, translation and rotation.

First we scale the fractal using (26) to ensure that at least for two subsequent nodes $a, b$ the absolute distance between these two nodes is equivalent. Since by (26) and (27) and the fact that both of these fractals are exact self-similar, we are sure that the distance between any subsequent pair of corresponding nodes on the two fractals is equivalent. By virtue of the fact that total angular change (27) between branch nodes is equivalent and that absolute distance between consecutive branch nodes is as well, we can rotate and translate one fractal such that symmetry lines between the branches overlap with the other. As shown in image 7 there remains a distance $d$ between corresponding branch points.

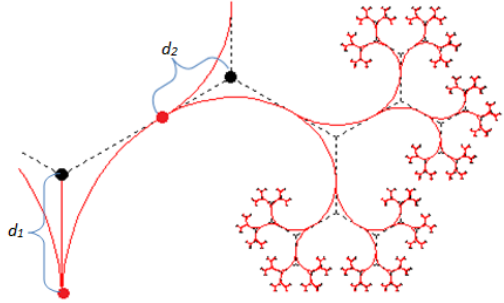

Image 7: Distance between corresponding nodes on two similar fractals

Due to the exact self-similar property (19) we know that

$$d_2 = d_1 \rho \text{ and } 0 < \rho < 1 \tag{28}$$

Since the branch points $s_i$ are equidistant (21) when $s \to \infty$, the node generations $i \to \infty$ and the distance $d$ between two corresponding nodes on fractals $\breve{p}_A(s)$ and $\breve{p}_B(s)$ is

$$d = \lim_{i \to \infty} d_1 \rho^i \tag{29}$$

Since $0 < \rho < 1$ it follows that $d = 0$. Hence the canopies are equivalent through scaling, translation and rotation. □

### 4.2.3 Branch point programming

#### *Binary trees and tree fractals*
In the previous example we stated that the branch points in $s$ were at equidistant intervals. Using Theorem 2, we can introduce branch points arbitrarily to create different trees. Below is a tree that has the same derivative coordinate functions as the fractal in Image 3 but with arbitrary branch points along $s$.

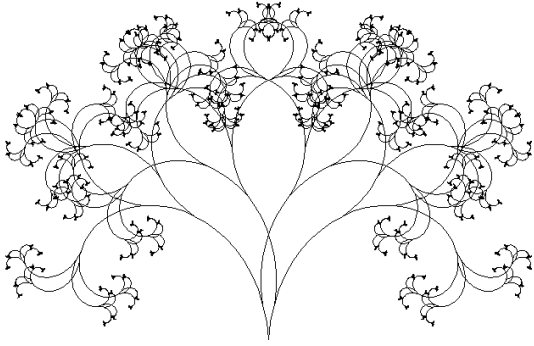 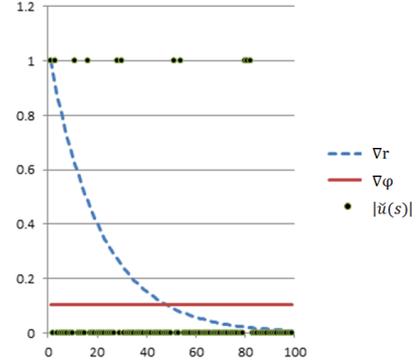

Image 8: Smooth fractal : $\Phi = \frac{\pi}{3}$ and $\nabla r = \left(\frac{2}{3}\right)^{s/\Delta s}$ 　　　Image 9: Corresponding $\nabla r$, $\nabla \varphi$ and $|\breve{u}(s)|$

It may appear that this tree has lost its self-similarity, however this branch point pattern may well repeat itself so the tree may still be qualitative fractal. This is why we will sometimes refer to such trees as fractal.

### *n-ary trees and tree fractals*

In practice, we do not have to limit ourselves to binary trees. We may create n-ary trees and fractals or we can create trees with different numbers of branches at different generations of branch points. The theorems 1,2 3 and their corollaries may be adapted to include these cases but we will not do so in this paper.

When implementing such mixed n-ary trees and fractals, which may be in higher dimensions the programming of branch points and their relationship with the number of branches, the appropriate derivative coordinate functions and dimensional direction a systematic approach is advised. For some applications where the branch geometry of primary interest an L-system grammar can be used very effectively [1].

Other applications, such as in engineering, a modular approach that focuses on sub trees may be more suitable.

### 4.2.4 Accessory functions

Theorems 1,2 and 3 are useful formulations for mathematical analysis. For applications in engineering they are useful for creating and optimizing tree and fractal structures or even architectures. In such practical environments we need to extend the model of derivative coordinate functions to include other features than a tree or fractal skeleton.

***Definition 4.1.*** *We define a set of derivative accessory functions $\nabla b_i$ for $c_i: \mathbb{R} \mapsto \mathbb{R}^n$, $i, n \in \mathbb{N}$ and $n > 0$ in which we include the derivative coordinate functions and a set of non-derivative accessory functions $b_j$ for $c_j: \mathbb{R} \mapsto \mathbb{R}^k$, $j, k \in \mathbb{N}$ and $k > 0$ to enhance the formulation of a tree or fractal defined in theorem 2. The set of derivative accessory functions is:*

$$C(s) = \{\nabla r, \nabla \varphi, \nabla b_0 \cdots \nabla b_n, c_0 \cdots c_k\} \tag{29}$$

*The accessory functions, together with the fractal function $\breve{p}(s)$, specify the enhanced fractal*

$$\breve{P}(s) = \{\breve{p}(s), C(s)\}. \tag{30}$$

### *Derivative vs non-derivative accessory functions*

Here we have defined derivative functions to ensure that they are dependent on their relative values, or history, rather than an absolute dependence on $s$. We may need this for line width, tube diameter and other relative accessory functions. We also have non-derivative functions that allow us to control absolute accessory functions. This may include color and material for example.

### *Static accessory functions*

The accessory functions are typically used to describe additional features. For graphical rendering this may be line width, color, transparency or any other aspect that is required. We may see accessory functions that define a type of material, a finish etc.

### *Interdependent accessory functions*

Accessory functions may be interdependent. If we have an accessory function that represents the water pressure in a system of fractal pipes, than it will depend on the size of the pipes which would be another accessory function.

### *Dynamic accessory functions*

In modelling we can also create accessory functions that are dynamical parameters such as temperature, flow or torque etc.

### *Sensory accessory functions*

Accessory functions may also depend on external parameters. One interesting example is an accessory function that measures the distance from the tree or fractal to a perimeter. This quantity can then be used to adjust the angular derivative coordinate function $\nabla\varphi$ such that it bends a branch away from or towards the perimeter.

### 4.2.5 Fractal engineering

We are now in a position to formally define a process for engineering trees and fractals. We aim to concatenate trees such that all of their properties, whether geometric or otherwise connect without undesired discontinuities.

**Theorem 5**. *Two trees $\breve{P}_1$ and $\breve{P}_2$ as defined by Theorem 2 may be concatenated to create a tree $\breve{P}$ with preservation of continuity by sequencing their accessory functions:*

$$\breve{P} = \breve{P}_1 \ll \breve{P}_2 \Rightarrow \tag{31}$$

$$\breve{P} = \{\breve{p}(s), C_1(s_1) + C_2(s_2)\} \tag{32}$$

*For the domain $s = \{s_1 \cup s_2\}$ is continuous and $\{s_1 \cap s_2\} = \emptyset$.*

*Proof.* It is sufficient to note that the derivative accessory functions, including the coordinate functions are specified as derivatives and that in evaluating their absolute function values, these derivative functions are integrated over their derivative variable $s$. As long as we ensure that the integration constants $h_1$ and $h_2$ for the definite integral of each derivate accessory function $\nabla b_1$ and $\nabla b_2$ is such that the resulting primitives $b_1(s) + h_1 = b_2(s) + h_1$, then concatenation will be smooth for each accessory function.

The non-derivative accessory functions do not required to connect smoothly since they represent absolute properties. □

## *Example of fractal engineering*

Starting with two well-known curves, a smooth Koch curve based on the straight lined fractal with parameters $\nabla r = (\alpha)^{s/\Delta s}$ where $\alpha = \dfrac{-1}{2\cos 144^o}$ and $\Phi = 144^o$ , a "golden fractal" by Tara D. Taylor[5]

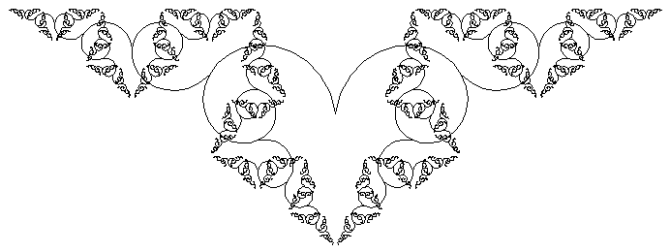

Image 10: Smooth Koch curve

and a smooth H-fractal

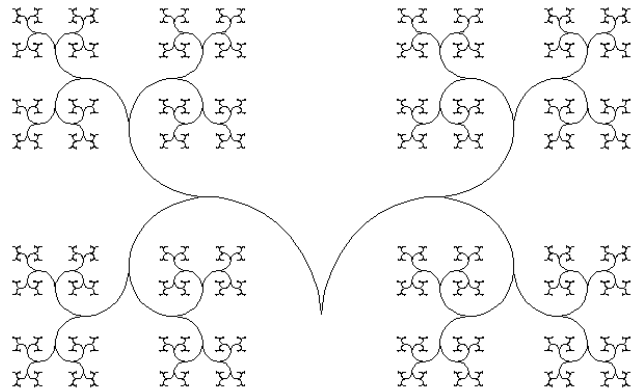

Image 11: Smooth H-fractal

we can concatenate the accessory functions:

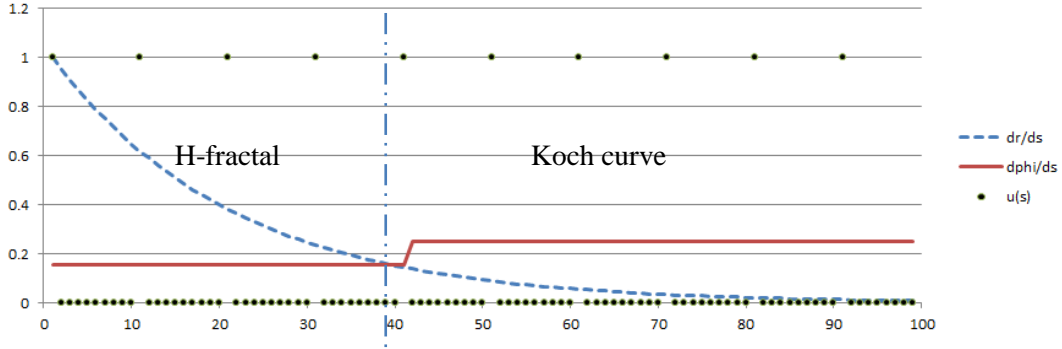

Image 11: concatenated derivative coordinate functions we obtain an engineered hybrid.

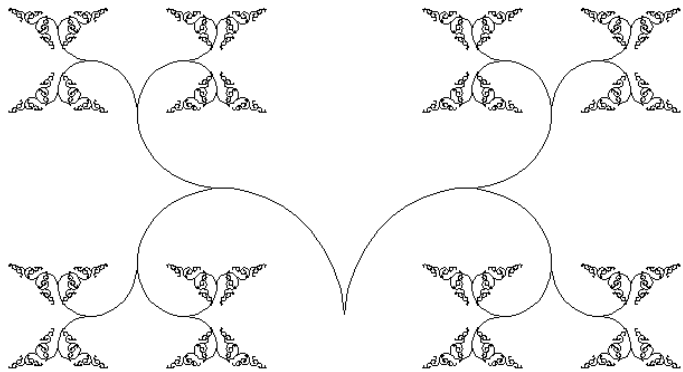

Image 12: Concatenated (hybrid) H-Koch fractal

### *Modular design in fractal engineering*

Theorem 4 opens up the door towards modular design in fractal engineering. All engineering disciplines have in common that systems and structures are created from subsystems and substructures. Componentization is at the core of engineering. Theorem 5 allows us to create trees by adding up sub-trees. There is no limit to this process we can use it to create low level structures that gradually become more useful as they combine into larger structures.

As yet, engineers have not found many applications for fractal structures, but this is likely a matter of maturity in design thinking. Below are some examples of engineered fractals and trees. The first example (Image 13) shows the use of sub systems, wheels and brackets. These subsystems in turn have subsystems in the form of cogs and spokes.

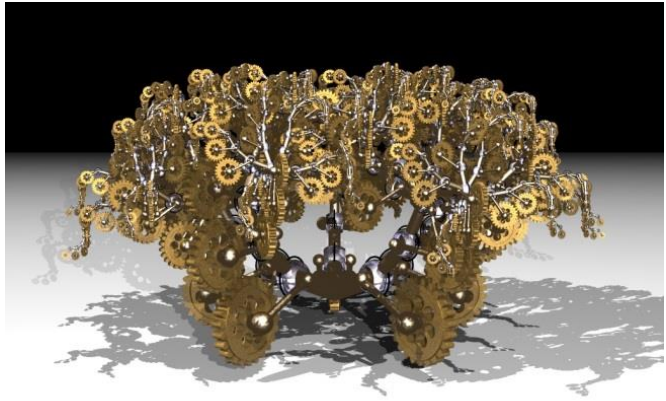

Image 13: Fractal system (sculpture) created from subsystems (wheels)

The second example of an engineered tree with a degree of fractal symmetry is shown below. The accessory functions of this object are a sequence of smooth functions, giving an organic impression. Note that the five forms share a root whose transparency accessory function renders it invisible. Although there is no use of subsystems in this tree, it is clear that elements of it could be reused as subsystems.

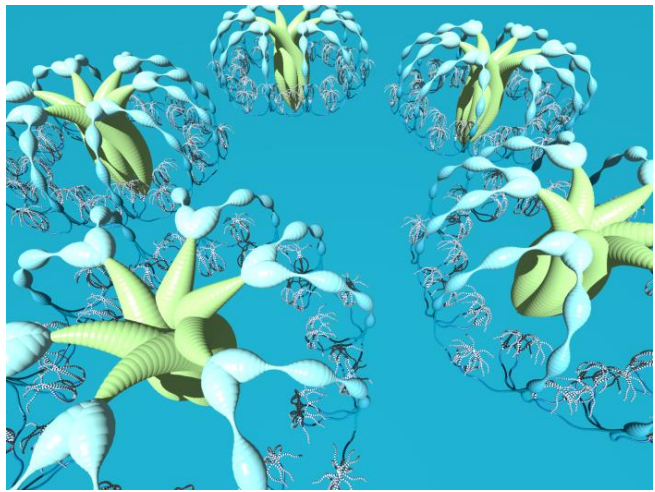

Image 14: Tree system (organic object) created from other sequenced forms (organs)

Such engineered fractals and trees created from components, suggest the opportunity for component libraries and tools for combining them. Genetic Fractals [8] experimented with a rudimentary library and programming language for creating such structures.

### 4.2.6 Applications

From an engineering perspective, fractals are a solution in search of a problem. Probably we will find many such problems in time. At present, the author has focused on two applications.

*Computer Graphics*

The use of derivative coordinates and the analytic fractal formulation is a powerful and simple approach to drawing and animating all manner of tree fractals in two or more dimensions. As with most fractals we see, there is an artistic incentive in this. The smooth nature of these fractals, combined with creativity allows us to explore new forms and ideas.

The image below shows a hybrid tree fractal that has 4 branches at the root, followed by 2 branches, 3 branches and a series of 6 branches. The accessory functions include color, branch width and transparency.

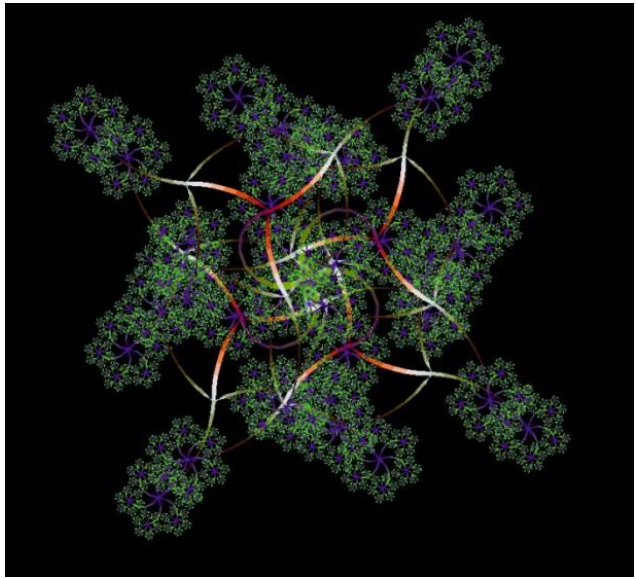

Image 15: Two dimensional hybrid tree fractal

The following image was adapted from an animation of a rotating four dimensional tree [9], i.e. it has branches into the second, third and fourth dimension consecutively.

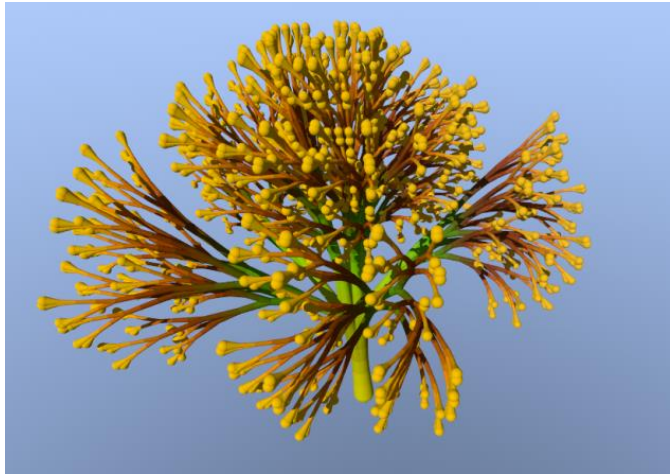

Image 16: Four dimensional tree fractal projected into three dimensional space

*3D Design and 3D Printing*

It is a small step to take three dimensional design into a 3D printing environment. There are some technical hurdles such as the need to convert the structure and its features into a triangulated mesh and a suitable file format such as STL. 3D printing itself has its limitations in printing resolution and choice and mixing of materials.

More importantly, and perhaps more fundamentally, the popularization of 3D printing has to overcome the 3D design obstacle. Traditional manufacturing relies on professional designers and sophisticated design tools for the creation of meaningful designs. The limitations of design tools suitable for non-professionals restrict 3D designs to simplistic and childlike quality. Although a lot of progress is made by the 3D design industry [10], non-professionals are not fully equipped yet to manipulate non-trivial geometrical forms in three dimensions on two dimensional screens. The modular approach of Theorem 4 introduces a one dimensional paradigm for 3D design. We need not concern ourselves with coordinates and rotations in three dimensions, we only have to decide how to sequence subcomponents (a one dimensional activity) and the tree formulation of Theorem 2 will take care of the complexities.

The example below shows a 3D printed tree fractal whose branches bend and curl [11].

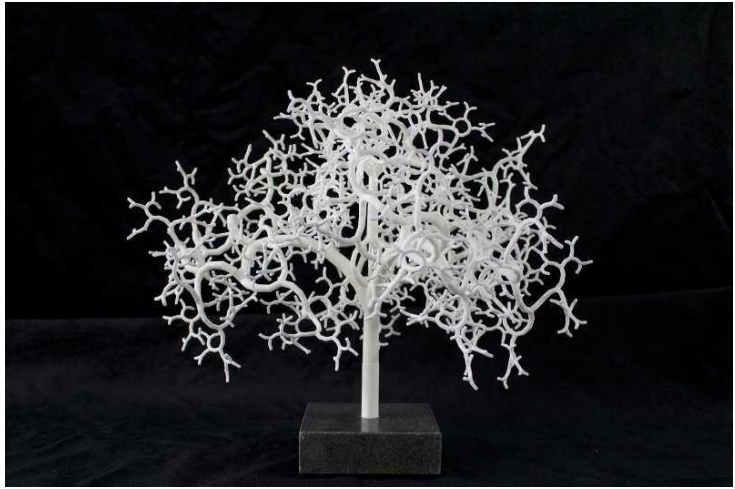

Image 17: 3D printed tree designed with the tree formulation of Theorem 3.

[image and 3D printing credit: PrintaBit.ch]

The last example below shows the wind blades of a hypothetical wind turbine where the wing ribs are a subsystem of the overall tree object. As the tree is evaluated, an algorithm calculates the triangular surface meshes required for 3D printers.

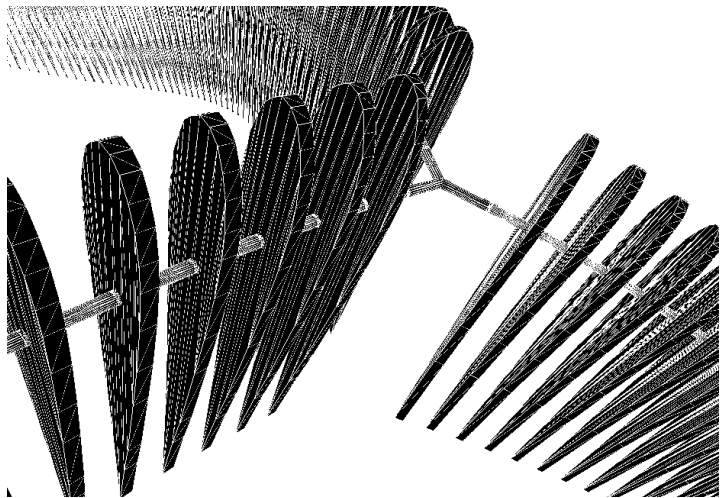

Image 18: design for wind blades with triangular mesh for 3D printing

# 5. Conclusion

## 5.1 Summary

We reviewed the analytic foundation of derivative coordinates in Theorem 1 that allows us to construct any path in n-dimensional space. We then created a model for branching using a multivalued unit function that allowed us to extend Theorem 1 to Theorem 2, i.e. an analytic formulation for binary symmetric trees, including smooth trees. Theorem 3 summarized the conditions for using this formulation for exact self-symmetric tree fractals.

Using these formulations we reviewed the computational advantages of using derivative coordinates and the tree and fractal formulations over traditional constructions using matrices. We then looked at a few examples of fractals created with this approach and demonstrated equivalence between straight lined tree fractals and fractals in Theorem 4.

We extended the model by introducing derivative accessory functions that allow us to model additional features that we find in engineered systems, such as materials, finishes etc. This led to Theorem 5 that allows us to do fractal engineering, i.e. building tree and fractal structures from components whilst ensuring structural integrity and continuity. Finally we reviewed a few applications of the theory in computer graphics and 3D printing.

## 5.2 Directions for further research

It would seem that there are a number of directions that the theory presented here can be further developed. Three areas appear to stand out.

### *Analysis of smooth trees and fractals*

Evidently, the ability to use analytical tools on the analytic formulation of trees and fractals should allow us to explore existing and new features of tree fractals analytically rather than numerically.

### *Development of complex fractal structures in engineering*

The use of fractal structures in engineering is still in its infancy. The analytic formulation of such structures will allow us to marry structure with structural analysis and dynamics.

### *Use of derivative coordinates in mathematics*

Derivative coordinates are independent from an absolute coordinate system. Beyond the use in trees and fractals, they may be used in any system with localized behavior.